\documentclass[aps]{revtex4}
\usepackage{graphicx}
\usepackage{bm}
\usepackage{amsmath}
\usepackage[T1]{fontenc}
\usepackage{mathrsfs,ae,dsfont}

\begin{document}

\title{Understanding quantum interference in General Nonlocality}
\author{Hai-Jun Wang}

\address{Center for Theoretical Physics and School of Physics, Jilin
University, Changchun 130012, China}

\begin{abstract}
In this paper we attempt to give a new understanding of quantum
double-slit interference of fermions in the framework of General
Nonlocality (GN) [J. Math. Phys. 49, 033513 (2008)] by studying
the self-(inter)action of matter wave. From the metric of the GN,
we derive a special formalism to interpret the interference
contrast when the self-action is perturbative. According to the
formalism, the characteristic of interference pattern is in
agreement with experiment qualitatively. As examples, we apply the
formalism to the cases governed by Schr\"odinger current and Dirac
current respectively, both of which are relevant to topology. The
gap between these two cases corresponds to the fermion magnetic
moment, which is possible to test in the near future. In addition,
a general interference formalism for both perturbative and
non-perturbative self-actions is presented. By analyzing the
general formalism we predict that in the nonperturbative limit
there is no interference at all. And by comparison with the
special formalism of Schr\"odinger current, the coupling strength
of self-action in the limit is found to be $\infty$. In the
perturbative case, the interference from self-action turns out to
be the same as that from standard approach of quantum theory. Then
comparing the corresponding coefficients quantitatively we
conclude that the coupling strength of self-action in this case
falls in the interval $[0,1]$.
\end{abstract}

\maketitle

\section{Introduction}

Up to date, the experiments on quantum double-slit interference/diffraction
have been realized with a series of single, separate fermions (electrons,
neutrons etc.) ~\cite{Claus}~\cite{Tono}~\cite{links}~\cite{Zeil}. A logical
inference is that the interference pattern is created by each single
particle interfering with itself. The given explanation is the Schr\"odinger
description \cite{Ahar} using the analogy to the optical Young's experiment
and Huygens-Fresnel principle. In this picture the fermion, one and the same
quantum entity, has to be viewed as a local particle in the source and on
the screen, but as a true matter wave for the propagating procedure in
between \cite{Nairz}. Accordingly, the interference fringe is believed to be
produced by the superposition of the waves from the two slits. It confuses
us by introducing classically split paths to \emph{a} matter wave. So far
quantum interference has been always explained by utilizing a somewhat
classical wave. In this paper we employ a realistic quantum nonlocal wave to
describe the whole interference procedure, and the very same quantum entity
can be viewed as a particle only when its charge(or one of other quantum
numbers) is detected. Our nonlocal method and the given explanation differ
in that the given explanation requires separate paths for a matter wave,
whereas our nonlocality needs only one spreading path.

~\\

The aim of this paper is to apply the theory of nonlocality \cite{Wanng} to
the nonlocal phenomenon---double-slit interference. There are other works ~%
\cite{Ahar}~\cite{Pope} specifying the relationship between the
nonlocality and the interference from different perspectives. In
the reference~\cite {Ahar} the authors attempt to link a set of
locally measurable operators (observables) with each nonlocal
procedure occurring between the source and the screen. These
operators are designed to be sensitive to the relative phase of
two slits, and to be involved in a kind of nonlocal interaction by
satisfying the nonlocal Heisenberg equation of motion. In this
sense the intermediate process of double-slit interference is
explained on a somewhat deterministic basis, so in principle, can
be measured too. Significantly, the authors put forth and explain
clearly the important fact that there are two kinds of
nonlocality: one is the well-known correlation born from the
violation of Bell-inequality concerning the Hilbert-space
structure; the other is relating to the typical issue of
double-slit experiment---in which the relative phase of two slits
cannot be observed locally---it is called a dynamical nonlocality.
The latter is the objective of this paper.

~\\

Different from other explanations to the quantum interference, in this work
we focus our attention on the performance of the matter wave rather than the
particles' position. We hold the point of view that the detection of the
position of a quantum particle relies on one kind of its charges/(or one of
its quantum numbers), whereas the detection of a wave usually depends on the
interferenc/defraction it produces. For a quantum entity, its wave is like a
wing around the charge(s), forming an inseparable entirety. In a sense it is
the stochastic behavior of the charge(s) in the wing that makes us realize
the nonlocality. When an interaction happens to the quantum particle, both
the ''wing'' and the charge(s)/(quantum numbers) are involved. In quantum
mechanics and conventional quantum field theory (CQFT), only the
charges/(quantum numbers) are mostly emphasized, whereas the low energy
limit--when the wave characteristic dominates the processes---has rarely
been considered. The two-slit interference is a paradigm to display wave's
nonlocality. And this kind of nonlocality cannot be completely described
solely by any conventional equation of motion, such as Schr\"odinger
equation or Dirac equation.

~\\

The starting point of this paper is the principle of the GN: A quantum wave
always remains an entirety (in a special complex reference frame, this
entirety is plane wave) regardless of interactions/observations (from the
point of view of quantum field, we regard any observation as a kind of
interaction). This principle has been applied to the metrics corresponding
to complex spaces to derive the equation of motion for fermions and the
field equation for bosons~\cite{Wanng}. Here we apply this principle to the
wave covering the two slits. Between the slits, namely the grating, is the
singularity mentioned in ~\cite{Wanng}. Since this wave keeps entirety in
the whole intermediate process, there is a moment when the wave breaks
itself at one side of the slits and simultaneously merges together at the
other side of the slits. At this very moment the wave forms a loop in the
3-dimension configuration space [fig.1c]. We describe this loop using the
metric of GN and the Action form in CQFT. The terminology of Action in CQFT
provides direct relevance to topology.

~\\

Since each intermediate process of interference involves only one fermion's
wave, we recognize that the two-slit interference phenomenon is
intrinsically nonlocal. Therefore the extended (nonlocal) wave must interact
with itself around the slits (the grating acts as a singularity). We name
such interaction as self-(inter)action henceforth and describe it by
fermion's current. This current-self-action is found to reside in the metric
of General Nonlocality (GN)\cite{Wanng}, which in the perturbative limit is
equivalent to a phase factor---an exponential of partial Action $S=\int %
\mathscr{L}\/d^4x$. The interaction term in the Lagrangian $\mathscr{L}$ now
becomes self-action in concept. The metric interpretation is assumed
available in our prescription even when the self-action is nonperturbative.
Our purpose is to understand the possible new effects suggested by the
self-action and non-perturbation. The self-action is new for conventional
quantum theory.

~\\

The remaining parts of the paper are arranged as follows. In section II we
derive a special formalism for interference phenomenon by starting from the
metric of GN and by considering only a classically thin "path" for a wave
[fig.2]. The formalism is relevant to the Action in CQFT in the case of
perturbative self-action. In Secs. III-IV the above special formalism is
applied to Schr\"odinger current and Dirac current, and the relevant
topology is discussed. Sec. V is dedicated to the general formalism for
interference where an extended "path" [fig.2](all the possibilities of
complex/configuration space) is involved. In Sec. VI we estimate the
strengths of self-action in different cases by analyzing relevant
experiments and comparing our results with conventional approaches of
quantum theory. Finally summary and remarks are presented.

\section{The special formalism for interference phenomenon}

The theory of GN~\cite{Wanng} states that a spatially-extended quantum wave
is a nonlocal entity as an entirety---''a point of complex space''. There
always exists a local complex reference frame in which the matter wave looks
like a plane wave (The wave is observed by another fermion or by itself), no
matter the existence of interactions/observations. Sometimes the
interaction/observation may be caused/performed by the change of
configurations. For example, while a wave passing through the double-slit
plane, the wave is to some extent interacting with the verges of the slits.
According to the principle of nonlocality, we infer that at any moment the
wave exists as an entirety, unlike the usual assumption that the wave is
split into two parts before they finally meet at a point on the screen. So
even if the splitting actually happens, it occurs after the wave fronts
having met and merged on the other side of the slit-plane [fig. 1]. We
interpret this kind of integrity (entirety) by transforming the integral
over the whole 4-dimensional space---used in metric of General
Nonlocality[see eq. (2.5)]---to a line integration along one spreading path
[see eq. (3.1)].

\begin{figure}[htbp]
\begin{center}
\includegraphics[width=5cm,height=8cm, angle=270]{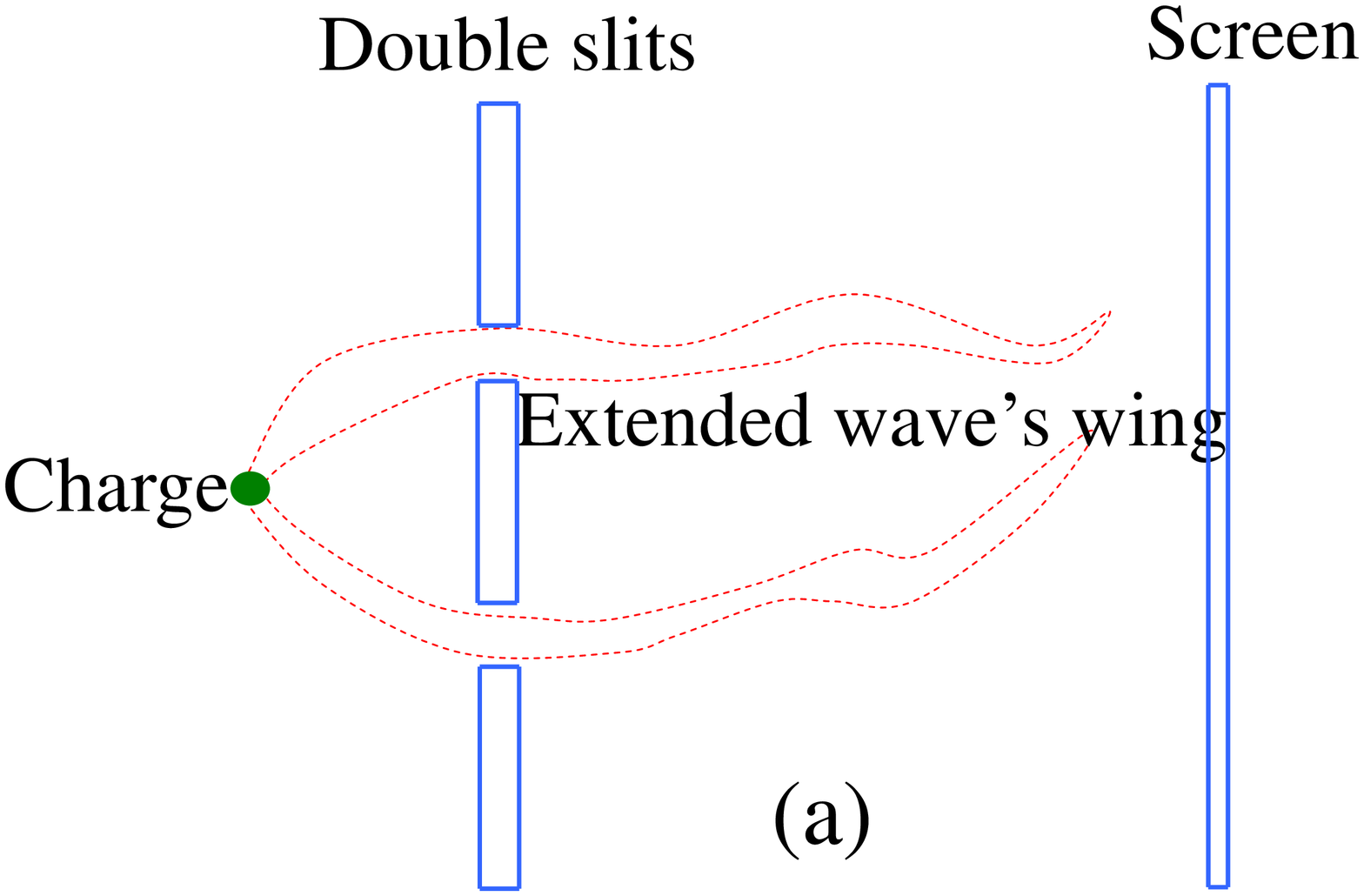} %
\includegraphics[width=5cm,height=8cm, angle=270]{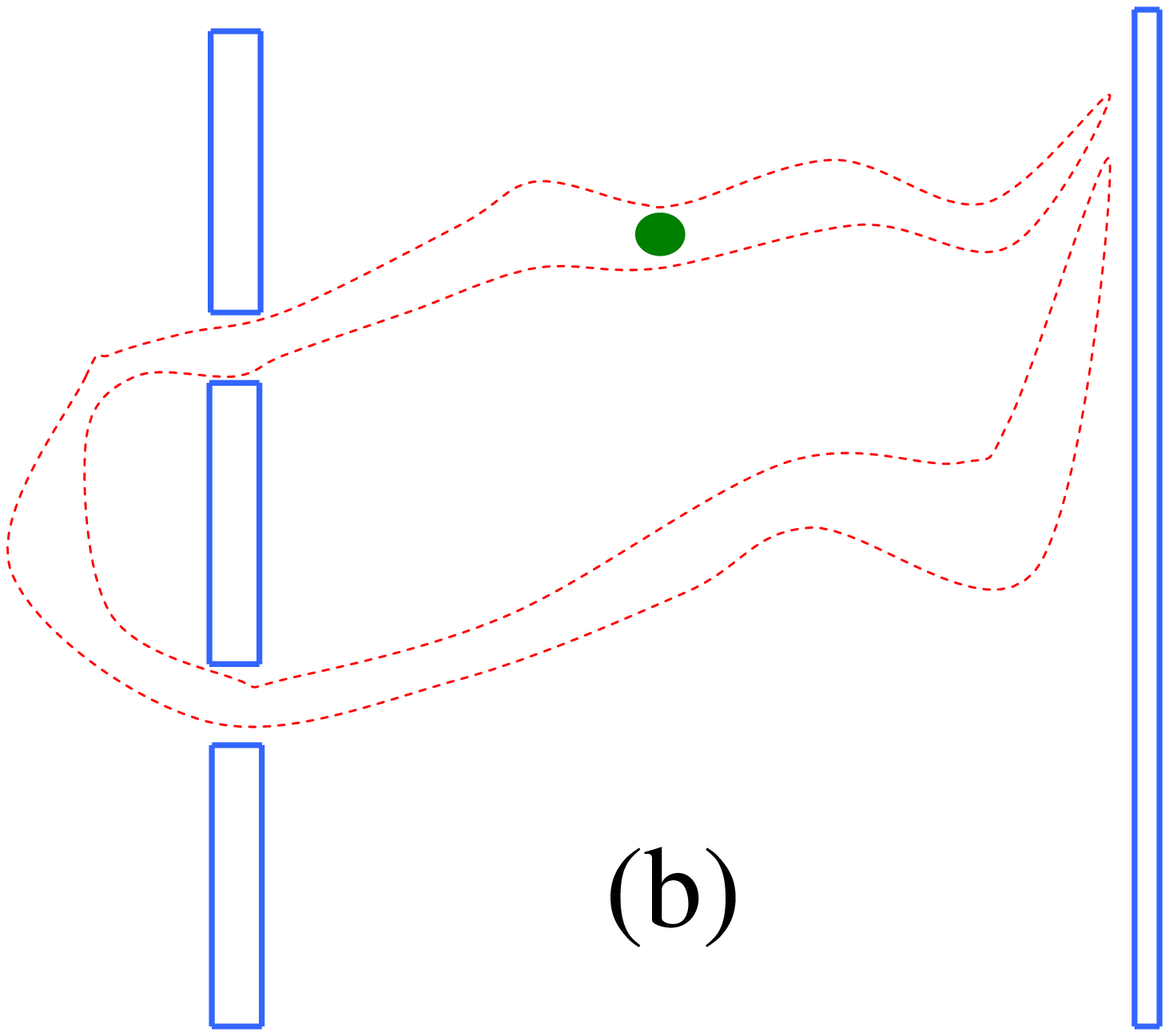} %
\includegraphics[width=5cm,height=8cm, angle=270]{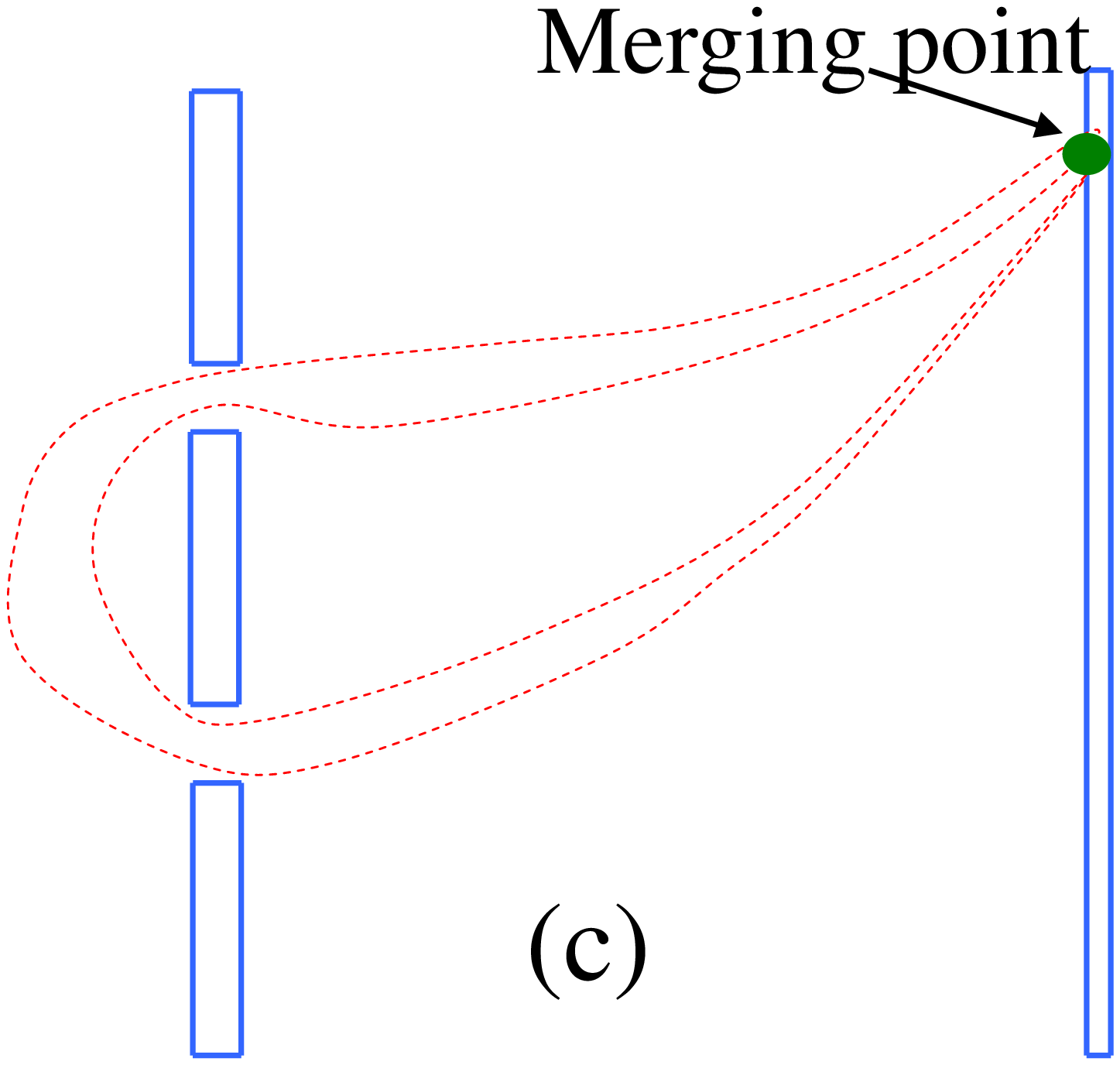} %
\includegraphics[width=5cm,height=8cm, angle=270]{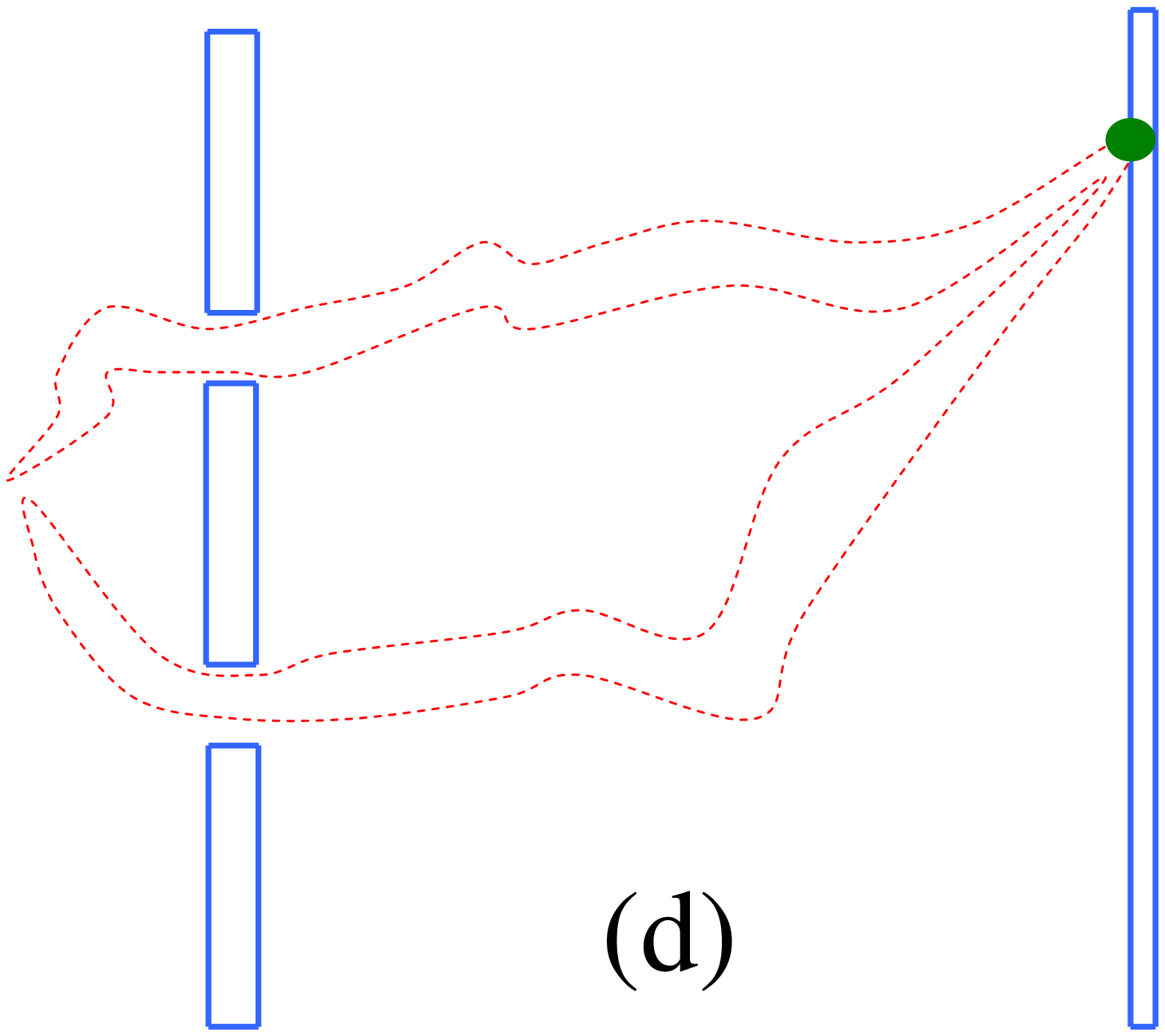}
\end{center}
\caption{We use scheme diagrams (1a)-(1d) to illustrate the procedure of a
matter wave evolving from one side of two-slit-barrier to another side,
until its two "ends" meet and merge on screen. Here the realistic wave's
scale is used, with only a few wavelengths involved.}
\label{fig 1}
\end{figure}

Accordingly, we start with the metric for the fermion fields,
\begin{equation}
G(\bar \psi ,\psi )=d\psi ^\alpha A_{\alpha \bar \beta }d\bar \psi ^\beta
\text{ ,}  \tag{2.1}
\end{equation}
here we will use the projected form of local wave function (in the paper ~%
\cite{Wanng} a projection was used to project the complex space to a flat
subset $d\psi ^\alpha \rightarrow \psi ^\alpha $, to enable the wave
function observable for a real observer). And we employ the approximate
metric tensor as
\begin{equation}
(A_{\alpha \bar \beta })_{4\times 4}=\gamma _0+\gamma _0\gamma ^\mu A_\mu
\text{ .}  \tag{2.2}
\end{equation}
Combining the above two equations yields
\begin{equation}
G(\bar \psi ,\psi )=\bar \psi \psi +j^\mu A_\mu \text{ ,}  \tag{2.3}
\end{equation}
$\tilde \psi =\{\psi ^1,\psi ^2,\psi ^3,\psi ^4\}$ is a
four-component spinor, and $j^\mu =\bar \psi \gamma ^\mu \psi $,
$\bar \psi =\psi ^{\dagger }\gamma _0$. This projected form
actually makes the metric tensor in a quasi-flat space---the usual
quantum-mechanical Hilbert space. For a function $\psi (\vec x,t)$
already projected onto flat space, in experiment however, it is
impossible to know its value at each point $\vec x$. So an
integral should be performed at the right hand side to sum over
all the possibilities (we have omitted this integral in
\cite{Wanng} for convenience, since performing the variation for
equation of motion is not affected by this omission), the
expression (2.3) alters to
\begin{equation}
G(\bar \psi ,\psi )=\int \bar \psi \psi d^3x+\int j^\mu A_\mu d^3x\text{ .}
\tag{2.4}
\end{equation}
As from eq. (8.8) to eq. (8.9) in \cite{Wanng}, by analyzing the dimension
we note that in the second term of the last equation (2.4) there lacks a
coordinate dimension in comparison with the first term. In the natural unit
where coupling constant $e=1$, if assuming the first term $\int \bar \psi
\psi d^3x=1$, then $\psi$ or $\bar\psi$ has the dimension of $[L^{-\frac{3}{2%
}}]$, here we use $[~~]$ to denote the dimension of a quantity and $[L]$ the
dimension of length. As for the second term, $[A_\mu]=[L^{-1}]$ and $%
[j^\mu]=[\bar \psi][\psi]=[L^{-3}]$ with obviously $[d^3x]=[L^{3}]$, so in
total the dimension of $\int j^\mu A_\mu d^3x$ is $[L^{-1}]$. This validates
the above judgement. The discrepancy of the dimensions is caused by the
projection aforementioned. To remedy the unequal dimensions of the two terms
and according to our experience of treating the Schr\"odinger equation with
a nonlocal interaction potential~\cite{WanngB}, we add a line integral to
the second term, with an imaginary number $i$ beforehand as did formerly,
\begin{equation}
G(\bar \psi ,\psi )=\int \bar \psi \psi d^3x+i\int j^\mu A_\mu d^4x\text{ .}
\tag{2.5}
\end{equation}

\begin{figure}[htbp]
\begin{center}
\includegraphics[width=5cm,height=8cm, angle=270]{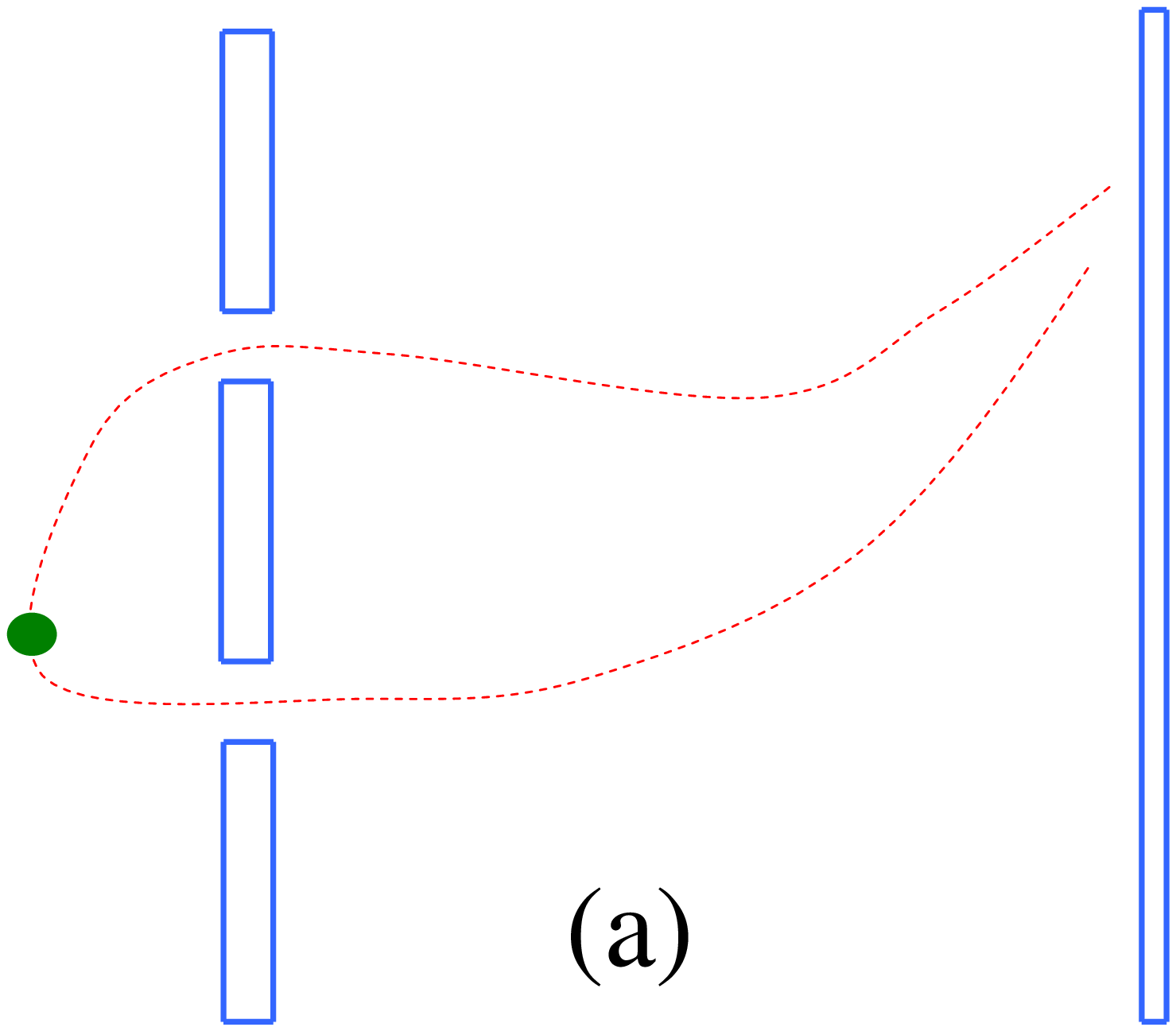} %
\includegraphics[width=5cm,height=8cm, angle=270]{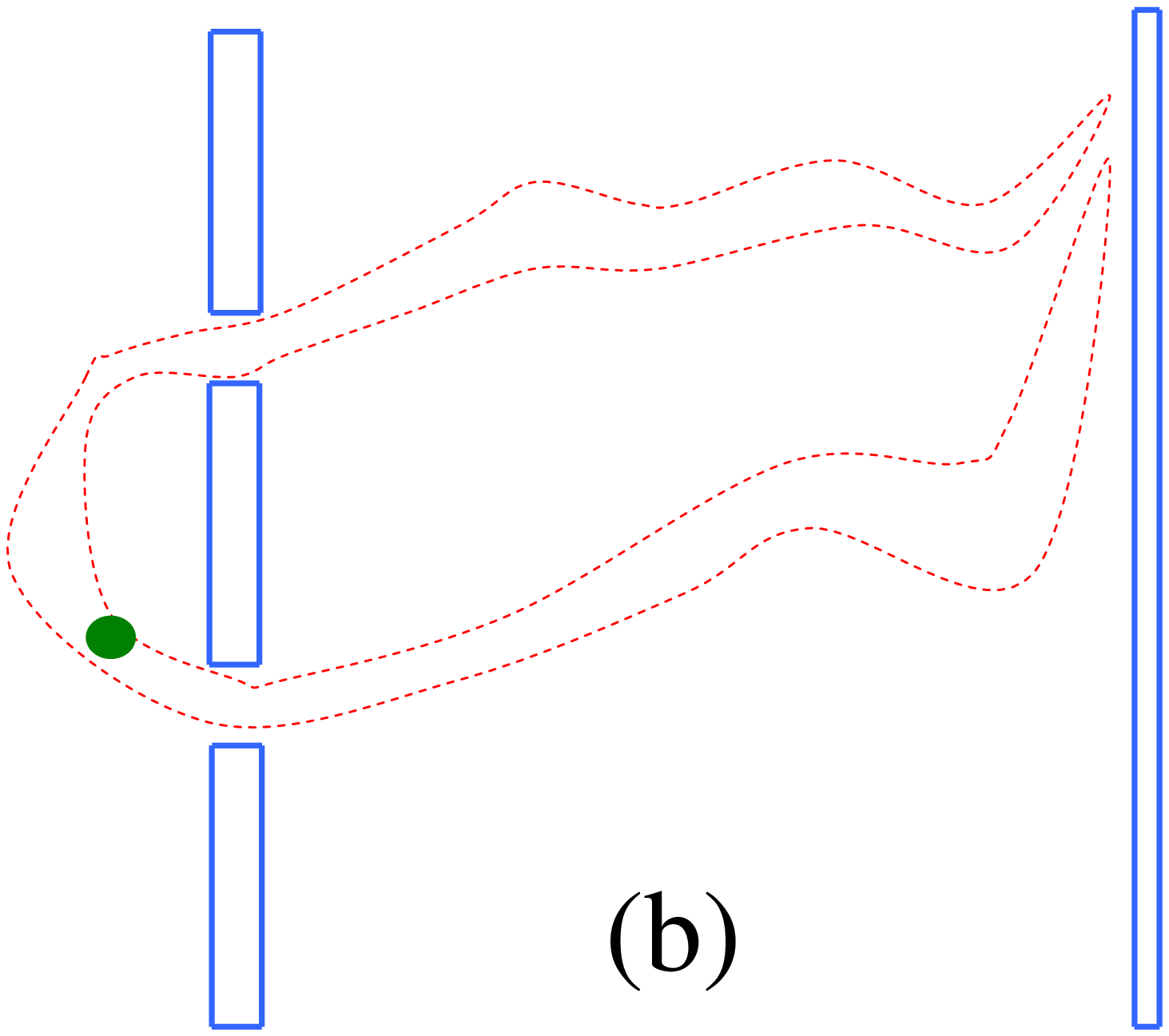}
\end{center}
\caption{Illustrating the meaning of "thin" path and "spreading" path, which
concern the different abstract volumes in the configuration space.}
\label{fig2}
\end{figure}

We recognize that while the (self-)interaction $A^\mu $ is very weak
(perturbative), the above expression associates with the phase factor $%
e^{iS} $ by
\begin{equation}
G(\bar \psi ,\psi )=1+i\int j^\mu A_\mu d^3x\,\text{d}t=1+i\int \mathscr{L}%
_I\/d^4x\sim e^{iS_I}\text{ ,}  \tag{2.6}
\end{equation}
in which $\int \psi ^{*}\psi d^3x=1$ is applied and the Action form $%
S_I=\int \mathscr{L}_I\/d^4x$ is understood, where $\mathscr{L}_I=j^\mu
A_\mu $ is interaction Lagrangian density (Later on we use the same form to
express the self-action.). From eq. (2.6) we realize that the physical
meaning of $G(\bar \psi ,\psi )$ (after projection) is linked with \emph{the
infinitesimally nonlocal evolution of the wave state}.

\begin{figure}[htbp]
\begin{center}
\includegraphics[width=8cm,height=12cm, angle=270]{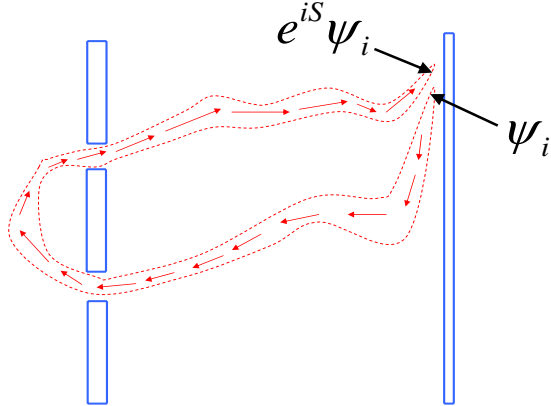}
\end{center}
\caption{At the merging point on the screen, $e^{iS}\psi _i$ meets $\psi _i$%
. }
\label{fig3}
\end{figure}

\begin{figure}[htbp]
\begin{center}
\includegraphics[width=8cm,height=12cm, angle=270]{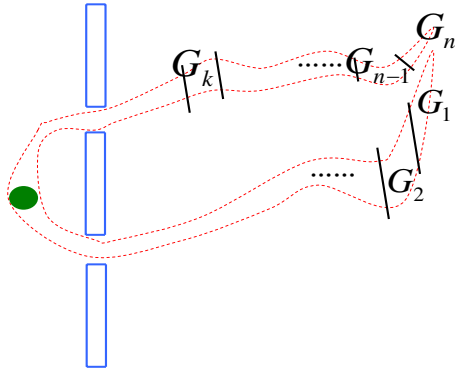}
\end{center}
\caption{Illustrating how the successive $G_n(\bar \psi ,\psi )^{\prime}s$
yield a loop expression.}
\label{fig4}
\end{figure}

According to Feynman's statement~\cite{FeynA}, a path contributes to
wave-function with a phase proportional to the action $S$,
\begin{equation}
\phi [x(t)]=\text{const }e^{i\,S[x(t)]/\hbar }\text{ ,}  \tag{2.7}
\end{equation}
hereafter we use natural unit by letting $\hbar =1$. Thus the wave $\psi
_i(x)$ evolving from one point $x$ infinitesimally to another point $x+dx$
along the same (thin) path may be expressed by the following form,
\begin{equation}
\psi _f(x+dx)=G(\bar \psi ,\psi )\,\psi _i=e^{iS_{\bigtriangleup }}\psi _i%
\text{ ,}  \tag{2.8}
\end{equation}
whereby the volume $d^3x\,$d$t$ in the integral $i\int j^\mu A_\mu
d^4x$ of $S$ is assumed to be infinitesimal, and the corresponding
$S$ is denoted by $S_{\bigtriangleup }$. So at the merging point
on the screen, the $e^{iS}\psi _i$ meets $\psi _i$ [fig.3], with
the merging wave as
\begin{equation}
\psi _{\text{merging}}=\psi _i+e^{iS}\psi _i\text{ ,}  \tag{2.9}
\end{equation}
where $S=\oint S_{\bigtriangleup }$ forming a loop [fig.4],
\begin{equation}
e^{iS}={\lim_{n\rightarrow \infty }}e^{iS_{\bigtriangleup _n}}\cdot \cdot
\cdot e^{iS_{\bigtriangleup _1}}={\lim_{n\rightarrow \infty }}G_n(\bar \psi
,\psi )\cdot \cdot \cdot G_1(\bar \psi ,\psi )\text{ .}  \tag{2.10}
\end{equation}

A more general form of eq. (2.10) will be found in the latter
part. This equation is just the special case with only one thin
path involved for a wave. The eq. (2.9) gives the probability of
finding the fermions by
\begin{equation}
\mid \psi _i+e^{iS}\psi _i\mid ^2=\mid e^{-iS/2}\psi _i+e^{iS/2}\psi _i\mid
^2=4\cos ^2\frac S2\mid \psi _i\mid ^2\text{ ,}  \tag{2.11}
\end{equation}
which is very the general Cosine form from experimental interference fringes
for fermions \cite{Tono}. It's known that if the Action form $S_I$ contains
a conventional kinetic part like $S_{\text{k}}=\int \mathscr{L}_{\text{k}%
}\/d^4x=\int \bar \psi (i\not \partial -m)\psi \/d^4x\sim \int \frac{\vec p^2%
}{2m}\/dt$ (the last step holds for low velocity, $\frac{\mid \vec
\upsilon \mid }c\sim 0$), the $\cos ^2\frac S2$ in eq. (2.11)
could coincide with conventional approach ~\cite{FeynA} \cite{Gli}
\cite{Bar}. We leave the discussion of this part implicit here. As
for the consistency of lacking this part in $S_I$, please refer to
the argument following eq. (5.7) and Sec.VI. In Secs. III and IV
we will show that it is almost identical to use Schr\"odinger
current or to use Dirac current in the Action form $S$.

\section{The Schr\"odinger current---its selfaction and topology}

In sense of nonlocality\cite{Wanng}, at each (arbitrary) moment
the wave exists as an entirety, not as usually assumed that the
wave has been split into two parts by slit-plane before they
finally meet at a point on the screen. Thus far we have admitted
that even if the splitting actually happen, they must occur when
the wave fronts having met and merged on the other side of the
slit-plane [fig.1d]. So we arrive at the conclusion that there is
at least one moment the wave shapes itself to form a loop in
three-dimension configuration space. This very loop yields the
following topological description.

We can go one step further from eq. (2.6) if the self-action has the form
like QED,
\begin{equation}
G(\bar \psi ,\psi )=e^{iS}=e^{i\int j_\mu A^\mu d^4x}=e^{iq\int ds_\mu A^\mu
}\text{ ,}  \tag{3.1}
\end{equation}
by using the relation
\begin{equation}
\int j_\mu A^\mu d^4x=\int \frac{\rho (x)ds_\mu }{dt}A^\mu d^4x=q\int ds_\mu
A^\mu \text{ ,}  \tag{3.2}
\end{equation}
where $q$ is the charge of the fermion. Please caution that $s_\mu $ is born
from a current and it is allowed to have some inner structure. If the
current is induced by an electron then $q=e=1$ in our default choice. For a
general current, $q$ may just be a constant. Moreover, there should exist
another charge $q^{\prime }$ in $A^\mu $, which will be used later.
Reiterating the argument at the beginning of this section, we note that the
summing up line-integral should be a loop form, which renders eq. (3.1) to
have the following form
\begin{equation}
e^{iS}={\lim_{n\rightarrow \infty }}G_n(\bar \psi ,\psi )\cdot \cdot \cdot
G_1(\bar \psi ,\psi )=e^{iq\oint ds_\mu A^\mu }\text{ .}  \tag{3.3}
\end{equation}

Since the intermediate process of each event in interference involves only
one fermion's wave, we can safely infer that the extended (nonlocal) wave
must interact with itself around the slits (the grating between the slits
forming a singularity). We will interpret this self-action in analogy to the
current-current coupling in weak interaction ~\cite{Fermi}, like the form $%
j_\mu j^\mu $ apart from a coupling constant. Accordingly the immediate
conclusion is
\[
A^\mu \sim j^\mu \text{ , for }A^\mu \text{ in eq.(3.3).}
\]
We prefer the current form $j^\mu =(\rho ,\mathbf{J})$ to appear in the line
integration eq. (3.3), since it has intriguing inner structures, and could
be a potential observable as well.

Now let's consider a type of self-action regardless of temporal component.
Since there is no instantaneous self-action relating to the local charge $q$
(As for the case of self-action, there is at most one point-wise charge
present, which is understood to be local and inseparable), we infer $A^0=0$.
However, since the 3-dimensional current $\mathbf{J}$ is intrinsically
nonlocal, its self-action is naturally allowed to occur. Under these
justifications, the eq. (3.3) reduces to
\begin{equation}
e^{iS}=e^{i\,q\oint d\mathbf{x}\cdot \mathbf{A}}\sim e^{i\,q\oint d\mathbf{x}%
\cdot \mathbf{J}}\text{ .}  \tag{3.4}
\end{equation}
The following Schr\"odinger current is our objective in this
section,
\begin{equation}
\mathbf{A\sim J}=-\frac i{2m}[\varphi ^{*}(x)\vec \nabla \varphi (x)-(\vec
\nabla \varphi ^{*}(x))\varphi (x)]\text{ ,}  \tag{3.5}
\end{equation}
in the Sec. IV we will show that Dirac current $j^\mu =\bar \psi
(x)\gamma _\mu \psi (x)$ behaves in a similar way. The current in
eq. (3.5) has an alternative form ~\cite{WanngA}
\begin{equation}
\begin{tabular}{l}
$\mathbf{J}=\frac \rho m\mathbf{\tilde J}\text{ }$, \\
$\tilde J_\mu =-\frac{\varphi ^{*}(x)\partial _\mu \varphi (x)-\varphi
(x)\partial _\mu \varphi ^{*}(x)}{2i\varphi ^{*}(x)\varphi (x)}\text{ , now }%
\mu =1,2,3$ , \\
$\rho =\varphi ^{*}(x)\varphi (x)$ .
\end{tabular}
\tag{3.6}
\end{equation}
Note that there is a dimension [$L^2$] difference between $\mathbf{J}$ and $%
\mathbf{\tilde J}$ (for [$\mathbf{J}$]=[$L^{-3}$], [$\mathbf{\tilde J}$]=[$%
L^{-1}$]), and likewise between $j^\mu $ and $A^\mu $, so we refine the
relation $A^\mu \sim j^\mu $ as $\mathbf{A}=\mathbf{\tilde J}$.
Consequently, the eq. (3.4) becomes

\begin{equation}
e^{iS}=e^{i\,q\,q^{\prime }\oint d\mathbf{x}\cdot \mathbf{\tilde J}}\text{ ,}
\tag{3.7}
\end{equation}
with $q^{\prime }$ also from the current $\mathbf{J}$.

To study the topology of the current, let's decompose the wave functions in
the above eq. (3.6) as follows
\begin{equation}
\begin{tabular}{l}
$\varphi (x)=\phi _1+i\phi _2\text{ ,}$ \\
$\varphi ^{*}(x)=\phi _1-i\phi _2\text{ ,}$%
\end{tabular}
\tag{3.8}
\end{equation}
and substituting them back to the eq. (3.6), one has
\begin{equation}
\tilde J_\mu =\frac{\phi _1\partial _\mu \phi _2-\phi _2\partial _\mu \phi _1%
}{\phi _1^2+\phi _2^2}\text{ ,}  \tag{3.9}
\end{equation}
furthermore making $n_1=\frac{\phi _1}{\sqrt{\phi _1^2+\phi _2^2}},n_2=\frac{%
\phi _2}{\sqrt{\phi _1^2+\phi _2^2}},$ yields
\begin{equation}
\tilde J_\mu =\varepsilon _{ab}n^a\partial _\mu n^b\text{ , }a\text{, }b=1,2%
\text{.}  \tag{3.10}
\end{equation}
According to Gauss-Bonnet formula in differential geometry ~\cite{Chern},
the integral of the above vector gives rise to an integer multiplied with
constant $2\pi $
\begin{equation}
\oint \tilde J_\mu dx^\mu =2\pi k_i,\text{ }k_i\text{ is an integer,}
\tag{3.11}
\end{equation}
substituting it into eq. (3.7), one obtains
\begin{equation}
e^{iS}=e^{i\,q\,q^{\prime }\oint d\mathbf{x}\cdot \mathbf{\tilde J}%
}=e^{i\,q\,q^{\prime }\,2\pi \,k_i}\text{ ,}  \tag{3.12}
\end{equation}
here the quantity $q\,q^{\prime }$ plays a role in expressing the strength
of the self-action, just as the coupling constants in gauge theory for
quantum fields. Substituting the form of $S$ in eq. (3.12) into eq. (2.11)
one finds that the width of the interference fringes is determined by the
quantity $q\,q^{\prime }$. In turn, the coupling strength $q\,q^{\prime }$
can also be deduced from the interference pattern. We will further
investigate the value of $q\,q^{\prime }$ in sect. VI.

\section{From Dirac current to Schr\"odinger current}

We have discussed the relationship between the current and its topological
properties starting from the simplest case of Schr\"odinger current. To make
our argument more general, we should guarantee the Dirac current includes
the Schr\"odinger current as a special case. In what follows we briefly
derive the Schr\"odinger current from Dirac current. The Dirac equation is
\begin{equation}
i\frac \partial {\partial t}\psi (x)=[\vec \alpha \cdot (-i\vec \nabla
-e\vec A)+eA_0+m\beta ]\psi (x)\text{ ,}  \tag{4.1}
\end{equation}
where $\alpha _\kappa =\left(
\begin{array}{cc}
0 & \sigma _\kappa \\
\sigma _\kappa & 0
\end{array}
\right) $ and $\beta =\left(
\begin{array}{cc}
1 & 0 \\
0 & -1
\end{array}
\right) $. From the equation one has obtained the conserved current as
follows ~\cite{Pes},
\begin{equation}
J_\mu =\bar \psi (x)\gamma _\mu \psi (x)=(\rho ,\mathbf{J})\text{ ,}
\tag{4.2}
\end{equation}
where $\gamma _0=\beta =\left(
\begin{array}{cc}
1 & 0 \\
0 & -1
\end{array}
\right) $ and $\gamma ^k=\beta \alpha _k=\left(
\begin{array}{cc}
0 & \sigma _\kappa \\
-\sigma _\kappa & 0
\end{array}
\right) $. Now we attempt to transform the above eq. (4.2) to Schr\"odinger
current form eq. (3.5).

To proceed, let's recall the large component method that we have used in
eqs. (9.13)\symbol{126}(9.16) of ~\cite{Wanng}. The eq. (9.14) is
\begin{equation}
\psi =\left(
\begin{array}{c}
\psi _1 \\
\psi _2 \\
\psi _3 \\
\psi _4
\end{array}
\right) =\left(
\begin{array}{c}
\psi _a \\
\psi _b
\end{array}
\right) \text{ ,}  \tag{4.3}
\end{equation}
with $\psi _a$ being the large component and $\psi _b$ the small component, $%
\psi _b$ can be interpreted as
\begin{equation}
\begin{tabular}{l}
$\psi _b=\frac{\vec \sigma \cdot (-i\vec \nabla -e\vec A)}{2\,m}\psi _a=%
\frac{\vec \sigma \cdot (\vec P-e\vec A)}{2\,m}\psi _a\text{ ,}$ \\
$\psi _b^{\dagger }=\psi _a^{\dagger }\frac{\vec \sigma \cdot (-i\stackrel{%
\leftarrow }{\nabla }-e\vec A)}{2\,m}=\psi _a^{\dagger }\frac{\vec \sigma
\cdot (-\stackrel{\leftarrow }{P}-e\vec A)}{2\,m}\text{ .}$%
\end{tabular}
\tag{4.4}
\end{equation}
Now we separately substitute the above approximation into eq.(4.2), to get
the approximate form of $\rho $ and $\mathbf{J}$. It yields
\begin{equation}
\rho =\psi ^{\dagger }\psi =(\psi _a^{\dagger },\psi _b^{\dagger })\left(
\begin{array}{c}
\psi _a \\
\psi _b
\end{array}
\right) \sim \psi _a^{\dagger }\psi _a\text{ ,}  \tag{4.5a}
\end{equation}
\begin{equation}
\begin{tabular}{l}
$\mathbf{J}=(\psi _a^{\dagger },\psi _b^{\dagger })\left(
\begin{array}{cc}
0 & \vec \sigma \\
\vec \sigma & 0
\end{array}
\right) \left(
\begin{array}{c}
\psi _a \\
\psi _b
\end{array}
\right) =\psi _b^{\dagger }\vec \sigma \psi _a+\psi _a^{\dagger }\vec \sigma
\psi _b$ \\
$=\frac{\vec \sigma \cdot (-\vec P-e\vec A)}{2\,m}\psi _a^{\dagger }\vec
\sigma \psi _a+\psi _a^{\dagger }\vec \sigma \frac{\vec \sigma \cdot (\vec
P-e\vec A)}{2\,m}\psi _a$ \\
$=\frac 1{2m}[\vec \sigma \cdot (-\vec P)\psi _a^{\dagger }\vec \sigma \psi
_a+\psi _a^{\dagger }\vec \sigma \vec \sigma \cdot \vec P\psi _a]-\frac
e{2m}[\psi _a^{\dagger }\vec A\cdot \vec \sigma \,\vec \sigma \psi _a+\psi
_a^{\dagger }\vec \sigma \,\vec A\cdot \vec \sigma \psi _a]\text{ ,}$%
\end{tabular}
\tag{4.5b}
\end{equation}
then employing the relation among the components of $\vec \sigma $,
\begin{equation}
\sigma ^j\sigma ^k=\delta ^{jk}+i\varepsilon ^{jkl}\sigma ^l\text{ ,}
\tag{4.6}
\end{equation}
we can obtain
\begin{equation}
\begin{tabular}{l}
$\vec \sigma \cdot (-\vec P)\psi _a^{\dagger }\vec \sigma \psi _a+\psi
_a^{\dagger }\vec \sigma \vec \sigma \cdot \vec P\psi _a$ \\
$=i[(\partial _j\psi _a^{\dagger })\sigma ^j\sigma ^k\psi _a-\psi
_a^{\dagger }\sigma ^k\sigma ^j\partial _j\psi _a]$ \\
$=i[(\partial _j\psi _a^{\dagger })(\delta ^{jk}+i\varepsilon ^{jkl}\sigma
^l)\psi _a-\psi _a^{\dagger }(\delta ^{kj}+i\varepsilon ^{kjl}\sigma
^l)\partial _j\psi _a]$ \\
$=-i[\psi _a^{\dagger }\vec \nabla \psi _a-(\vec \nabla \psi _a^{\dagger
})\psi _a]+[(\vec \nabla \psi _a^{\dagger })\times \vec \sigma \psi _a-\psi
_a^{\dagger }\vec \sigma \times (\vec \nabla \psi _a)]\text{ ,}$%
\end{tabular}
\tag{4.7a}
\end{equation}
\begin{equation}
\begin{tabular}{l}
$\psi _a^{\dagger }\vec A\cdot \vec \sigma \,\vec \sigma \psi _a+\psi
_a^{\dagger }\vec \sigma \,\vec A\cdot \vec \sigma \psi _a$ \\
$=\psi _a^{\dagger }A_j\sigma ^j\sigma ^k\psi _a+\psi _a^{\dagger }\sigma
^kA_j\sigma ^j\psi _a$ \\
$=A_j[\psi _a^{\dagger }(\sigma ^j\sigma ^k+\sigma ^k\sigma ^j)\psi _a]=0%
\text{ .}$%
\end{tabular}
\tag{4.7b}
\end{equation}
So the eq.(4.5b) becomes
\begin{equation}
\mathbf{J}=-\frac i{2m}[\psi _a^{\dagger }\vec \nabla \psi _a-(\vec \nabla
\psi _a^{\dagger })\psi _a]+\frac 1{2m}[(\vec \nabla \psi _a^{\dagger
})\times \vec \sigma \psi _a-\psi _a^{\dagger }\vec \sigma \times (\vec
\nabla \psi _a)]\text{ ,}  \tag{4.8}
\end{equation}
in which $\psi _a^{\dagger }$ is a $2\times 1$ matrix. The above equation
has already taken the shape of Schr\"odinger current, apart from an
additional term $(\vec \nabla \psi _a^{*})\times \vec \sigma \psi _a-\psi
_a^{*}\vec \sigma \times (\vec \nabla \psi _a)$ (we name this part as
\textit{spin current}) due to the inclusion of spin $\vec \sigma $. For the
interference pattern induced by single fermions, the fringes should have an
additional shift induced by the self-action of spin $\vec \sigma $, just
like the contribution of $\vec \sigma _i\cdot \vec \sigma _j$ in potential
model~\cite{WanngB}.

The \textit{spin current} can be further transformed into
\begin{equation}
\mathbf{J}_\sigma =\frac 1{2m}[(\vec \nabla \psi _a^{*})\times \vec \sigma
\psi _a-\psi _a^{*}\vec \sigma \times (\vec \nabla \psi _a)]=\frac 1{2m}\vec
\nabla \times (\psi _a^{*}\vec \sigma \psi _a)\text{ ,}  \tag{4.9}
\end{equation}
which just coincides with the Ampere law expressed in Maxwell equations
\begin{equation}
\vec \nabla \times \vec B=\vec j+\frac{\partial \vec E}{\partial t}\text{ .}
\tag{4.10}
\end{equation}
The eq. (4.9) cannot be further written as in eq. (3.10) which is
convenient for directly discussing topology. This term however,
formally relates to topology in a manner named Aharonov-Casher
(AC)~\cite{AC} effect, though the physics here is irrelevant to AC
effect. Likewise the eq. (3.4) is formally similar to
Aharonov-Bohm (AB) effect ~\cite{AB}. From eq. (4.10) we realize
that the term $\psi _a^{*}\vec \sigma \psi _a$ is just the
magnetic moment of a fermion. While the interference happens, it
surely accumulates its effect on the interference pattern due to
eq. (3.5). Based on the experience of comparing AC
effect~\cite{ACE} and AB effect~\cite{ABE}, this shift from $\psi
_a^{*}\vec \sigma \psi _a$ should be less than 0.1 percent of the
original. Thus in nonlocality, the strength of coupling $\vec
\sigma _i\cdot \vec \sigma _j$ may have the order of magnitude
$1/1000$ as the self-action in eq.(3.12). We expect this tiny
interference effect could be tested by experiment, e.g. via
comparing the interference patterns induced by single protons and
single neutrons, if possible.

\section{The general formalism for quantum interference}

This section is dedicated to giving a \textit{general} expression of eq.
(2.10), ${\lim_{n\rightarrow \infty }}G_n(\bar \psi ,\psi )\cdot \cdot \cdot
G_1(\bar \psi ,\psi )$, which helps us to get the interference terms like
those in eq. (2.9), $\psi _{\text{merging}}=\psi _i+(sth)\psi _i$. Here the
\textit{general} form means that we should take into account the diffusive
and extended (spreading) characteristic of the wave. This characteristic is
interpreted in Feynman path integral by performing the integral with respect
to the measure $[D\bar \psi ][D\psi ]$, which we will use for reference. So,
the expression we will treat is
\begin{equation}
{\lim_{n\rightarrow \infty }}\int [D\bar \psi ][D\psi ]G_n(\bar \psi ,\psi
)\cdot \cdot \cdot G_1(\bar \psi ,\psi )\text{ .}  \tag{5.1}
\end{equation}
The calculation doesn't belong to Feynman path integral unless the
self-action is perturbative, though it shares the same form as
Feynman integral ~\cite{Explain}. Here we first review the
treating of perturbative case, then cope with the non-perturbative
case.

Let's rewrite the integral eq. (5.1) explicitly in a piece-wise manner by
substituting the eq. (2.1),
\begin{equation}
{\lim_{n\rightarrow \infty }}\frac 1N\int [D\bar \psi _n][D\psi _n]\cdot
\cdot \cdot [D\bar \psi _1][D\psi _1](d\psi _n^\alpha A_{\alpha \bar \beta
}d\bar \psi _n^\beta )\cdot \cdot \cdot (d\psi _1^\alpha A_{\alpha \bar
\beta }d\bar \psi _1^\beta )\text{ ,}  \tag{5.2}
\end{equation}
in which we use the convenient formalism before \textbf{projection}---like
in Ref.~\cite{Wanng}, each $d\psi _n$ standing for the local wave function
for an infinitesimal short scale in complex space, and $d\bar \psi _nd\psi
_n $ representing the local probability of finding the particle. We add a
normalization factor $\frac 1N$ to normalize the volume of measure space $%
\int [D\bar \psi _n][D\psi _n]\cdot \cdot \cdot [D\bar \psi _1][D\psi
_1]=N\int d\bar \eta _nd\eta _n\cdot \cdot \cdot d\bar \eta _1d\eta _1$, $%
\eta ^{\prime }$s are Grassmann variables.

\subsection{Perturbative Case}

First let's treat eq. (5.2) in perturbative case, whereby the local metric $%
G_n(\bar \psi ,\psi )$ has the expanding form as~\cite{Wanng}[see sec.VI],
\begin{equation}
G_n(\bar \psi ,\psi )=d\psi _n^\alpha A_{\alpha \bar \beta }d\bar \psi
_n^\beta =d\psi _n^\alpha A_{\alpha \bar \beta }d\bar \psi _n^\beta \approx
d\bar \psi _n(1+\gamma ^\mu A_\mu )d\psi _n\text{ ,}  \tag{5.3}
\end{equation}
in which $d\bar \psi =d\psi \,\gamma _0$ and each component $A_\mu $ is
perturbatively small. As a result, the wave function $\psi $ only feels tiny
disturbance. So if $\psi $ is a kind of plane wave $e^{ikx}$, then $d\psi $
must be of the same kind except for a tiny phase shift $\delta $ which, in
the following eq. (5.4) is cancelled out due to the exponentials $e^{i\delta
}$ and $e^{-i\delta }$ in $\psi $ and $\bar \psi $. Approximately, we admit $%
\psi \sim d\psi _n$ (we identify this with a projection, $d\psi
_n\rightarrow \psi $). Thus in perturbative case,
\begin{equation}
G_n(\bar \psi ,\psi )\sim \bar \psi (1+\gamma ^\mu A_\mu )\psi \text{ .}
\tag{5.4}
\end{equation}
We should note that for different piece $G_n(\bar \psi ,\psi )$ the
self-action form $A_\mu $ may be varied, at least coordinate-dependent after
projection. After this projection, and concerning the integral in eq. (2.5),
one has
\begin{equation}
G_n(\bar \psi ,\psi )\sim \int \bar \psi \psi d^3x+i\int j^\mu A_\mu
d^4x\sim 1+i\int \mathscr{L}_I\/d^4x\sim e^{iS_I}\text{ .}  \tag{5.5}
\end{equation}
By this treatment, the eq. (5.2) turns out to be the similar form like
generating functional based on Feynman-path-integral,
\begin{equation}
{\lim_{n\rightarrow \infty }}\frac 1N\int [D\bar \psi _n][D\psi
_n]\cdot \cdot \cdot [D\bar \psi _1][D\psi _1](d\psi _n^\alpha
A_{\alpha \bar \beta }d\bar \psi _n^\beta )\cdot \cdot \cdot
(d\psi _1^\alpha A_{\alpha \bar \beta }d\bar \psi _1^\beta )\sim
\frac 1N\int [D\bar \psi ][D\psi ]e^{i\sum_nS_I^n}\text{ ,}
\tag{5.6a}
\end{equation}
the final result with reducing degrees of freedom is ~\cite{Pes}
\begin{equation}
\frac 1N\int [D\bar \psi ][D\psi ]e^{i\sum_nS_I^n}=\frac 1N\int
[D\bar \psi ][D\psi ]e^{i\int \bar \psi \not A\psi d^4x}=\det \not
A\text{ .}  \tag{5.6b}
\end{equation}
In the physical sense of standard generating functional, the
integral eq. (5.6b) should be equal to 1 if we add a kinetic term
to the phase of integrand, $i.e.$
\begin{equation}
\frac 1N\int [D\bar \psi ][D\psi ]e^{iS_{\text{k}}-i\int \bar \psi \not
A\psi d^4x}=\frac 1N\int [D\bar \psi ][D\psi ]e^{i\int \bar \psi (i\not
\partial -\not A-m)\psi d^4x}=1  \tag{5.7}
\end{equation}
This equation gives us the direct justification that the interference
induced from self-action $\int \bar \psi \not A\psi d^4x$ and that from
kinetic term $\int \bar \psi (i\not \partial -m)\psi d^4x\sim \int \frac{%
\vec p^2}{2m}\/dt$ \cite{Bar} should be mutually cancelled in eq.
(5.7). Combining them together yields the disappearance of the
interference. To put further, the interference patterns derived
from these two ways are identified with each other (This statement
to some extent implied by the aforementioned characteristic of the
self-action eq. (3.5), $A^\mu \sim j^\mu $). In the Sec. VI we
will use this property to discuss the value of $q\/q^{\prime }$ in
perturbative case.

\subsection{Nonperturbative Case}

In the cases of nonperturbatively strong self-action, the exponential form
of eq. (2.6) no longer holds since the second term of eq. (2.5) may be very
large. Therefore the factor of the second term by no means has the form $%
e^{iS}$. One may attempt to separate the metric as in eq. (5.3), in which
one part is perturbative. While $A_\mu $ is not perturbatively small, one
cannot do that even indirectly. We have to start from eq. (5.1) again, and
find a way out by repeating the principle of nonlocality: There always
exists a local complex reference frame in which a quantum wave looks like a
plane wave, no matter the existence of interactions/observations. The
principle ensures that there is a local transformation corresponding to the
local metric $G_n(\bar \psi ,\psi )$ to make the interaction formally
vanish, as follows,
\begin{equation}
G_n(\bar \psi ,\psi )=d\psi _n^\alpha A_{\alpha \bar \beta }d\bar \psi
_n^\beta =d\bar \psi _n(1+\gamma ^\mu A_\mu )d\psi _n\stackrel{\text{a local
transformation}}{\longrightarrow }d\bar \psi _n^{\prime }d\psi _n^{\prime }%
\text{ ,}  \tag{5.8}
\end{equation}
in which a transformation $J_n(J_n^{\dagger })$ between $d\psi _n^{\prime
}(d\psi _n^{\prime \dagger })$ and $d\psi _n(d\psi _n^{\dagger })$ is
implicit,
\begin{equation}
\begin{tabular}{l}
$d\psi _n=J_nd\psi _n^{\prime }\text{ ,}$ \\
$d\psi _n^{\dagger }=d\psi _n^{\prime \dagger }J_n^{\dagger }\text{ .}$%
\end{tabular}
\tag{5.9}
\end{equation}
The transformation $J_n(J_n^{\dagger })$, as suggested by GN, belongs to a
group larger than $SL(2,\not C)$ or $SU(N)$, for instance, the group $GL(4,%
\not C)$ or $U(N)$. For the time being, we mainly consider the case that $%
J_n(J_n^{\dagger })$ belongs to the larger group $GL(4,\not C)$. Writing eq.
(5.9) explicitly,
\begin{equation}
d\psi _n^{\dagger }\gamma _0(1+\gamma ^\mu A_\mu )d\psi _n=d\psi _n^{\prime
\dagger }J_n^{\dagger }\gamma _0(1+\gamma ^\mu A_\mu )J_nd\psi _n^{\prime }%
\text{ ,}  \tag{5.10}
\end{equation}
and comparing it with eq. (5.8), we have
\begin{equation}
J_n^{\dagger }\gamma _0(1+\gamma ^\mu A_\mu )J_n=\gamma _0\text{ .}
\tag{5.11}
\end{equation}

However, the transformation (5.9) will not lead to the variations of the
corresponding measure,
\begin{equation}
\begin{tabular}{l}
$\lbrack D\psi _n]=[D\psi _n^{\prime }]\text{ ,}$ \\
$\lbrack D\psi _n^{\dagger }]=[D\psi _n^{\prime \dagger }]\text{ ,}$%
\end{tabular}
\tag{5.12}
\end{equation}
since the degree of freedom of the measure should be free of the
transformation of $J_n$s$(J_n^{\dagger }s)$. If the freedom of measure $%
[D\psi _n]([D\psi _n^{\dagger }])$ belongs to $SL(2,\not C)$, then a
Jacobian determinant would appear for $[D\bar \psi _n][D\psi _n]$ after the
transformation, which is definitely divergent~\cite{Bao}~\cite{Kaku} and
brings about obstacles for further studies. In the above expression we
simplify the denotation by using $\psi _n(\psi _n^{\prime })$ instead of $%
d\psi _n(d\psi _n^{\prime })$ while they appear together with $D$. Now the
eq. (5.2) yields
\begin{equation}
{\lim_{n\rightarrow \infty }}\frac 1N\int [D\bar \psi _n^{\prime }][D\psi
_n^{\prime }]\cdot \cdot \cdot [D\bar \psi _1^{\prime }][D\psi _1^{\prime
}](d\bar \psi _n^{\prime }d\psi _n^{\prime })\cdot \cdot \cdot (d\bar \psi
_1^{\prime }d\psi _1^{\prime })\text{ .}  \tag{5.13}
\end{equation}

To evaluate the eq. (5.13) further, we have to evaluate the integrand $%
(d\bar \psi _n^{\prime }d\psi _n^{\prime })$ for each piece. Then we find it
approaching zero $d\bar \psi _n^{\prime }d\psi _n^{\prime }\stackrel{%
n\rightarrow \infty }{\rightarrow }0$ since, as aforementioned, $d\bar \psi
_n^{\prime }d\psi _n^{\prime }$ stands for the probability that the particle
exists in that piece, and in very low energy limit---the energy scale on
which nonperturbation appears---the particle tends to live at everywhere (as
an entirety) rather than a definite small regime in phase/configuration
space. So it yields
\begin{equation}
{\lim_{n\rightarrow \infty }}\text{ }\frac 1N\int [D\bar \psi _n^{\prime
}][D\psi _n^{\prime }]\cdot \cdot \cdot [D\bar \psi _1^{\prime }][D\psi
_1^{\prime }](d\bar \psi _n^{\prime }d\psi _n^{\prime })\cdot \cdot \cdot
(d\bar \psi _1^{\prime }d\psi _1^{\prime })=0\text{ .}  \tag{5.14}
\end{equation}
Inverting the above steps backward to eq. (5.1) and then to eq. (2.9), we
conclude that in the case of strong self-action, there is no interference.
Involving the unitary group $U(n)$ in $J_n$s$(J_n^{\dagger }s)$ the
conclusion would remain the same, with the similar logic.

If we write the integral as the form in eq. (5.6b), $\frac 1N\int
[D\bar \psi ][D\psi ]e^{i\int \bar \psi \not A\psi d^4x}$, then
the straightforward conclusion is
\begin{equation}
\det \not A=0  \tag{5.15}
\end{equation}
which could be a criterion when the self-action is (or tends to be)
nonperturbative. Coincidentally in GN, this condition is used to judge if a
system is of bound state. This means that for quantum systems with just
bound state, there is no observable interference phenomenon.

The two cases considered above involve all the possible paths in
phase/configuration space, namely the diffusive and extended
characteristics of the wave. However, while describing the
interference, usually we are not concerned about this
respect------one thin path is enough. In the Secs. III and IV we
have just done that for perturbative cases. Involving all the
paths in this section conduces to getting the following criterion,
\begin{equation}
\det \not A=0\text{ for non-perturbative self-action.}  \tag{5.16}
\end{equation}
The above conclusion for strong interaction applies to single path
equally, since for strong self-action we note that the result of
eq. (5.14) depends only on its integrand.

One may wonder what if we apply the eqs. (5.8)\symbol{126}(5.13) to
perturbative case. It seems that the same conclusion (5.14) would appear.
However, if one is aware that the condition $\psi \sim d\psi _n$ cannot be
met in nonperturbative case and the $(d\bar \psi _n^{\prime }d\psi
_n^{\prime })\rightarrow 0$ is not pertinent in perturbative case, then one
understands these two cases are essentially different. For the very reason
there is no necessity to put the perturbative system into a larger
configuration space governed by the group $GL(4,\not C)$ or $U(N)$.

\section{Analyses on experiments and determination of the strength $%
q\,q^{\prime }$}

\subsection{Analyses on experiments and evaluation of $q^{\prime }$}

It is necessary for us to estimate what on earth the value of
coupling strength $q\,q^{\prime }$ is. To this end, we may resort
to the earlier experiments \cite{Claus}~\cite{Tono} which
confirmed the existence of the interference of fermions' matter
wave. We find however, these experiments are demonstration
experiments, though they actually unveiled the wave aspects of
micro-massive-particle. To demonstrate the interference effect,
the experiments take at least two essential steps for
magnification. At the beginning, the electron is accelerated
through 50kV thus gains a corresponding deBroglie wavelength about
0.05$\stackrel{\circ }{A}$. With such a small wavelength electron
cannot interfere with itself when passing by macro-apparatus (the
macro obstacle acting as grating of double-slit). So firstly, a
\emph{transverse electric field} $V_a$ (henceforth we use
denotation $V_{\text{T}}$) is applied to act as the first biprism
to drag the wavelets from parallel ways to meet and to interfere
[fig. 2 of \cite {Tono}]. And in formulae the transverse field
induces a transverse momentum responsible for the interference
term~\cite{Tono}. By the above technical trick, the interference
fringe is already produced, but still too small to be observed
directly. So secondly, it must be enlarged by a named projector
lenses. Finally the pattern can be viewed by eyes on the
fluorescent screen. In summary, the scales---such as the
wavelength $\lambda $ of matter wave, the grating space, and the
spacing of interference pattern etc., are all experimentally
varied and don't reflect the true states of fermions located at
solid material or those flying towards or outwards an atomic
structure with pores, for instance an electron scattered off
atomic nucleus. So the data from those demonstration experiments
can be referenced but cannot be applied directly to our cases in
Secs. II-V

~\\

\begin{figure}[htbp]
\begin{center}
\includegraphics[width=8cm,height=12cm, angle=90]{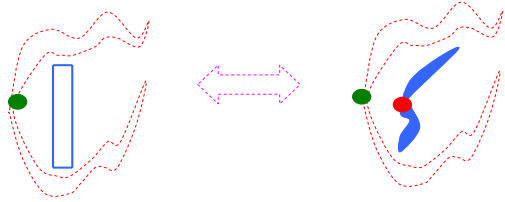}
\end{center}
\caption{To Illustrate schematically how we identify a quantum entity (a
particle with its wing---the wave) with the grating in double-slit
experiment.}
\label{fig5}
\end{figure}

The aforementioned transverse field, on one hand acts as a biprism of
focusing the wave to produce interference. On the other hand if we assume
the transverse field is induced by a source, then we note that in formalism
it provides another charge $\,q^{\prime }$. So in the eq. (6) of \cite{Tono}
there appears $qq^{\prime }$, which already exists in our formulae. This
observation suggests that the use of the transverse field in experiments is
consistent with our self-action assumption. Hence to some extent transverse
field also models the self-action. Now let's explain how a true
micro-process of matter-wave interference happens. As for an appropriately
arranged scattering processes, such as an electron scattering off a proton
in atomic nucleus, the electron's wavelength could be of the order of
magnitude of the scale of proton's radius, so that the proton simultaneously
acts as an grating of electron. While the electron heads on to the proton,
its wave will surround the grating (proton) and interact with itself [fig.
5]. This self-action is manifested by coupling strength $qq^{\prime }$.
However, this true interference and self-action cannot be observed directly
by our apparatus because of the very small scale. The true interfering
procedure seems more like a \emph{gedanken} experiment. We hope future
studies could give clues to directly observe the effect of the true
interference from scattering experiments.

~\\

Now we explain how we reference the experiments and obtain the estimation of
the coupling strength $qq^{\prime }$ in perturbative limit. In experiments~%
\cite{Tono}, about $V_{\text{T}}=10~Volt$ transverse potential difference is
used to change the directions of two wavelets, and produces a transverse
momentum $k_{\text{T}}$ responsible for deriving the interference fringe
(Here we omit the mechanism for deriving relationship between $V_{\text{T}}$
and $k_{\text{T}}$: $k_{\text{T}}\propto \frac{V_{\text{T}}}{\upsilon _{%
\text{L}}}$, $\upsilon _{\text{L}}$ is the longitudinal velocity of the
electron. For more details please refer to equations in \cite{Tono}).
Corresponding to the $k_{\text{T}} $ the transverse wavelength is $\lambda _{%
\text{T}}=\frac{2\pi }{k_{\text{T}} }\sim 500\stackrel{\circ }{A}$. We are
not concerned about the longitudinal wave effect for it takes such small
wavelength $0.05\stackrel{\circ }{A}$ that approximately it has no effect on
the interference, especially when we write the final wavefunction like $\psi
(x,z)=e^{ik_zz}\,(e^{ik_xx}+e^{-ik_xx})$. In our cases, the matter wave
interferes with itself and correspondingly the ''transverse field'' is
provided by itself. In order to compare the self-action with the above
experiment value quantitatively we propose that an incident electron heading
on a proton [fig. 5], with kinetic energy similar as the ground state of
hydrogen. We use hydrogen ground-state as paradigm and employ the Virial
theorem $2\langle T\rangle =-\langle V\rangle $ ($\langle \ \rangle $means
expectation value under a state) to evaluate the electric potential
difference (volt) for self-action. $E_0=-13.6eV$, using $\langle T\rangle
_0+\langle V\rangle _0=E_0$ leads to $\langle T\rangle _0=13.6eV$ and $%
\langle V\rangle _0=-27.2~eV$. Subsequently, one has $\lambda _{\text{T}%
}=\frac hp=h/\sqrt{2mE_k}\stackrel{here}{=}h/\sqrt{2m\langle T\rangle _0}%
\sim 3$ $\stackrel{\circ }{A}$ and $\upsilon _{\text{L}}=\frac{\lambda
~\langle T\rangle _0}h\sim 10^{-2}c$. According to the experience on
describing hydrogen, the electron with such kinetic energy tends to hold
itself around nucleon with average radius of the wavelength $\lambda $. In
such a manner, one end of the wave can feel the other. Self-action truly
happens. On such scale around the proton, the electron can provide for
itself the potential energy at most $\mid \langle V\rangle _0\mid =27.2~eV$.
And according to the requirement of the experiment $\lambda _{\text{T}}=%
\frac{2\pi }{k_{\text{T}}}$ and $k_{\text{T}}\sim \frac{V_{\text{T}}}{%
\upsilon _{\text{L}}}$~\cite{Ratio}, we take $V_{\text{T}}\sim 20~Volt$ to
fit $\lambda _{\text{T}}\sim $3 $\stackrel{\circ }{A}$. That means if we
take $q=e$, then we should take $q^{\prime }\sim 1.8\,e $ to divide $\mid
\langle V\rangle _0\mid =27.2~eV$ to get $V_{\text{T}}$. We conclude that $%
q\ q^{\prime }$ has the order of magnitude of $1$, but not necessarily equal
to 1.

~\\
\begin{figure}[htbp]
\begin{center}
\includegraphics[width=8cm,height=12cm, angle=90]{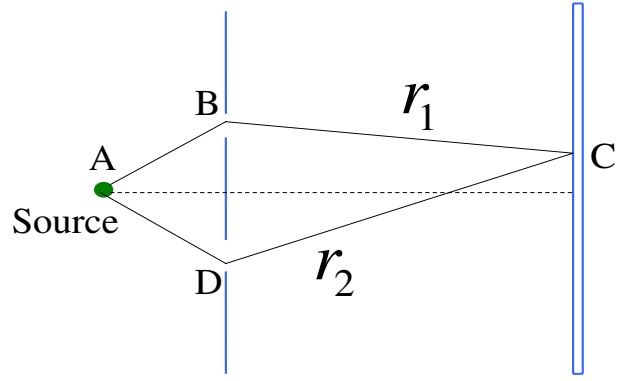}
\end{center}
\caption{To show the standard explanation on double-slit experiment in
quantum theory. $\triangle r=r_1-r_2$ denotes the difference of the paths
from two slits to a point on screen.}
\label{fig6}
\end{figure}

\subsection{Comparison with conventional approaches and determination of $%
q^{\prime }$}

Another way to determine the coupling strength $qq^{\prime }$ in the
perturbative limit is to compare the results from eqs. (2.11), (3.12), (3.4)
with standard approaches of quantum theory. We find surprisingly that $%
q^{\prime }$ is not a constant any longer in this way. As for the
conventional/standard approaches, we note that using Feynman Path Integral~%
\cite{Bar} and using Huygens-Fresnel principle [Fig. 6] yield the same
interference expression---proportional to $\cos ^2\theta =\cos ^2(\frac{2\pi
}\lambda \triangle r)=\cos ^2(\frac p\hbar \triangle r)$, where $\triangle r$
is the difference of paths from two slits to a point on screen $\triangle
r=r_1-r_2$. The expression can be derived from eqs. (11) (14) and (17) of
Ref.~\cite{Bar}. Referring to Fig. 6 we may write the difference $\triangle
r $ as follows
\begin{equation}
\triangle r=r_1-r_2=\widehat{ABC}-\widehat{ADC}=\int_{ABC}d\mathbf{x}%
-\int_{ADC}d\mathbf{x}=\oint_{ABCDA}d\mathbf{x}\text{ ,}  \tag{6.1}
\end{equation}
here we use $\widehat{ABC}$ to denote a path, and the subscripts of integral
signs denote the same thing. With the eq. (6.1) and $\cos ^2(\frac p\hbar
\triangle r)=\cos ^2\theta =\mid \frac 12(e^{i\theta /2}+e^{-i\theta
/2})\mid ^2=\mid \frac 12(1+e^{i\theta })\mid ^2$, and assuming the module
of momentum $p=\mid \mathbf{p}\mid $ is approximately a constant, $\cos
^2(\frac p\hbar \triangle r)$ can be formally written in the style of eq.
(2.11), with $e^{i\theta }$ as
\begin{equation}
e^{i\theta }=e^{i\oint \mathbf{p}\cdot d\mathbf{x}}\text{ ,}  \tag{6.2}
\end{equation}
the loop integral is along the path $\widehat{ABCDA}$, from now on the
natural unit is understood when $\hbar $ is omitted.

~\\

So far we have two ways to perform the comparison. One way is to
compare eq. (3.4), $i.e.$, $e^{i\,q\oint d\mathbf{x}\cdot
\mathbf{A}}$ ($q^{\prime }$ absorbed in $\mathbf{A}$) with eq.
(6.2), $i.e.$, $e^{i\oint \mathbf{p}\cdot d\mathbf{x}}$. Another
way is to compare $\cos ^2(qq^{\prime }2\pi k_i)$ [derived from
(2.11), (3.12)] with $\cos ^2(\frac p\hbar \triangle r)$.
Obviously the two ways are identified. If we take the former way,
then we go back to eq. (5.7) with another obvious form
\begin{equation}
e^{i\oint (\mathbf{p}-q\mathbf{A})\cdot d\mathbf{x}}=1\text{ .}  \tag{6.3}
\end{equation}
Eqs. (6.3),(5.7) show the consistency of our previous arguments---
the lack of kinetic part in the Action form of eq. (2.11) yields
the equivalence of our approach and conventional approaches in
perturbative case. However we cannot directly extract the value of
$qq^{\prime }$ from eq. (6.3). On the other hand, if we take the
latter way, then the straightforward conclusion is that the value
$q\/q^{\prime }$ depends on $\frac{\triangle r}\lambda $, since
$\frac p\hbar \triangle r=\frac{2\pi }\lambda \triangle r$. This
indicates that if the incident momentum (thus $\lambda $) is given
and $q=e=1 $, then $q^{\prime }$ is not a constant but varies with
$\triangle r$. For example, for the first maximum when $\theta
=2\pi $ and $\triangle r=r_1-r_2=\lambda $, from eq. (6.3) and
$\cos ^2(\frac p\hbar \triangle r)=\cos ^2(qq^{\prime }2\pi k_i)$
we can infer the value $A=\frac h{\triangle r}=p$, in such cases
$A$ gets the full value of $p$ , correspondingly $q^{\prime }=1$
(Here $q^{\prime }$ is just the ratio of modules $A=\mid
\mathbf{A}\mid $ and $p=\mid \mathbf{p}\mid $, not involved
their vector characteristic.). Mostly the value of $A$ is just a part of $p$%
. In this sense, the corresponding strength of $\mathbf{A}$, i.e., $%
q^{\prime }$, has its value in the interval $[0,1]$. Obviously the
self-action $A_\mu $ is path-dependent and not a conservative potential. The
path-dependence stems from the attempt to describe $A_\mu $---which is
intrinsically nonlocal---by using point-wise (particle) method. $A_\mu $
only shows its definite properties in nonlocal phenomenon when global
measurements are performed. In any local detection $A_\mu $ may display its
uncertainty, and here the value of $\mathbf{A}$ varies with the paths'
direction.

~\\

Now one may wonder what the relationship of the above self-action $A_\mu $
and momentum $p_\mu $ is. Our explanation is that for (self-)interaction $%
A_\mu $, in Dirac equation $(\not p-\not A)\psi =m\psi $ or in Schr\"odinger
equation $i\hbar \frac{\partial \psi }{\partial t}=[\frac{(\mathbf{p}-%
\mathbf{A})^2}{2m}+V(\vec r)]\psi $ (with charge $q^{\prime }$ already
absorbed in $\not A$ or $\mathbf{A}$), $p^\mu $ and $A^\mu $ ($\mathbf{p}$
and $\mathbf{A}$) are on the same foot. $A_\mu $ can be born from $p_\mu $,
and vice versa (This statement is to some extent implied by the
aforementioned characteristic of the self-action, eq. (3.5) $A^\mu \sim
j^\mu $). Since $\mathbf{A}$ is a vector, to speak $A$ as a part of $p$ may
induce confusion, for example possibly $\mid \mathbf{p-A}\mid >\mid \mathbf{p%
}\mid $. Whereas on the other hand, eq. (6.3) can reduce the complexity by
just performing an integral. As for the coupling strength $q^{\prime }$ we
don't consider such complexity and just use the results after the line
integral has been performed. At least this statement pertains the quantum
interference. In Introduction we have asserted that the wave of quantum
entity can be viewed as its wing. Now we may question what is on earth in
the wing (to speak equally, in the wave). Through the above analysis, we now
have confidence to conclude: in the wing, it is just the self-action
potential $A_\mu $, or the momentum $p_\mu $, or their mixture. When
interactions/observations occur, $A_\mu $ is responsible for interfering
(interacting) with itself and exhibiting its wave aspect, while $p_\mu $
plays the role of ''flying'' and defining paths of the particle. They
transfer to each other constantly and may not be separated physically. This
picture deserves further study.

~\\

For the nonperturbative limit, we have eq. (5.16): $\det \not
A=0$, which corresponds to $\frac 1N\int [D\bar \psi ][D\psi
]e^{i\int \bar \psi \not A\psi d^4x}=0$ and thus $e^{i\int \bar
\psi \not A\psi d^4x}=0$ for quantum interference. If assuming the
eq. (3.12) is also available to nonperturbative cases, we conclude
that only when $q^{\prime }q=-\infty $ does $\det \not A=0$ hold.
This indicates that if $q=e=1$ then $q^{\prime }$ takes the value
$\infty $, consistent with the word ''nonperturbative''. Thus far
we have certified the conclusions from eqs. (2.9), (3.12) and
(5.16) and ref. \cite{Wanng} are consistent for nonperturbative
limit.

\section{Summary and Remarks}

Throughout this whole paper we are pursuing a new understanding of quantum
interference based on the perspective of matter wave's self-action. By the
special formalism of the interference we find out that the interference
pattern has the common form of being the superposition of Cosine functions,
in accordance with the experimental results from single electrons. By
comparing the phase factors induced by Schr\"odinger current and Dirac
current, it follows that the fermions with and without a large magnetic
moment would give rise to different interference fringes due to a tiny
spin-current effect, which awaits for test by experiment. By analyzing the
general formalism of the interference, we predict that there is no
interference phenomenon when the self-action is nonperturbative. In
particular, we understand that the self-action $A_\mu $ can be born from
momentum $p_\mu $ by comparing the general formalism in perturbative limit
with conventional approaches of quantum theory. The strength of self-action
is found to fall in the interval $[0,1]$ by such comparisons.

~\\

A novel aspect of this paper is the formalism we have followed. From the
metric form $G(\bar \psi ,\psi )$ and its physical meaning---associated with
the infinitesimally nonlocal evolution of the wave state, we can derive the
formalism analogous to Feynman path-integral to interpret the interference
contrast (for this analogy we have borrowed some treating methods from CQFT
whenever possible). However we note from the relation $G(\bar \psi ,\psi
)\sim e^{iS_I} $ that it is not the true Feynman path integral since the
explicit form of $S_I$ is not the true Action. And the operators' form $%
j^\mu A_\mu $ in $S_I$ is more like the Hamiltonian for $S-$matrix. So the
understanding of this phase factor $e^{iS_I}$ becomes in a dilemma from the
perspective of CQFT. On one hand if we view it as Feynman path integral, $%
S_I $ lacks a kinetic integral $S_{\text{k}}=\int \bar \psi (i\not \partial
-m)\psi \/d^4x\sim \int \frac{\vec p^2}{2m}\/dt$. On the other hand if we
regard it as the $S-$matrix in scattering, it lacks the legs to define its
external propagators ~\cite{Ryder}.

~\\

The above discrepancy---lacking a kinetic term---originates from the
difference of the geometric method we have employed and the Lagrangian
method the CQFT used. In differential geometry, by construction it is not
concerned about the motion of the points (point in real space or complex
space), but about how the space (or the manifold) is curved. Once the
curvature is defined and fixed by $R=0$, the motion of the point (to speak
physically, the particle) is totally determined by the geodesic equation.
Similar arguments apply equally to General Relativity. Whereas in the
Lagrangian method, it works in a way of balancing the kinetic energy and the
interaction potential by $\delta \mathscr{L}=0$. So the kinetic energy has
to be present in Lagrangian. In our theoretical frame, the discussions are
dominated by the properties of space, say, the wave-function $\psi $%
---space. Whereas in CQFT, the discussions are dominated by the properties
of operators. In CQFT there are quantization processes (producing
operators), which is irrelevant in our theory. Summarizing from Sec.V and
Sec.VI we note that that discrepancy not only lays no obstacle for us to
discuss physics, but also drives us farther in obtaining the strength of
self-action.

~~\\

Future researches along this line fall into five respects: Firstly we may
extend present results to multi-slit and many-body systems. Secondly some
other kinds of important self-action should also be examined, for instance,
the axil vector form $\gamma _\mu \gamma _5$. Thirdly, so far the
interference phenomenon in crossover regime between perturbation and
nonperturbation remains unknown in the frame of GN. In addition, we may
apply the physical picture or even the conclusions to scattering cases if we
make in mind the grating between slits shrink to a quantum entity as shown
in Fig.5. Finally, the self-action itself also deserves further
investigation.

~\\

\section*{Acknowledgement}

I am grateful to Prof. H. Z. Zhang for raising his question on Nonlocality,
which became one of the motivations of this research. Also thanks a lot to
Dr. P. M. Zhang, Dr. D. J. Jia, Prof. A. D. Bao, Prof. S. H. Chen, and Prof.
W T Geng for many heuristic discussions. This work is supported in part by
Fundamental Research Funds for the Central Universities.


\begin{thebibliography}{99}
\bibitem{Explain}  At the starting point of Feynman integral, the nonlocal
characteristic has been partially included by equally summing over all the
paths. But this is ruined by the mostly used Gaussian integral and the least
phase method.
\bibitem{Ratio}  Phenomenologically, we keep the ratio between $k_{\text{T}}$
and $\frac{V_{\text{T}}}{\upsilon _{\text{L}}}$ unchanged in experiment~\cite
{Tono} and our estimation.
\end{thebibliography}
\end{document}